# Thermodynamic Analysis of Snowball Earth Hysteresis Experiment: Efficiency, Entropy Production, and Irreversibility

Valerio Lucarini [Email: v.lucarini@reading.ac.uk]

Department of Meteorology, University of Reading, Earley Gate, PO Box 243, Reading, RG6 6BB (UK)

Department of Mathematics, University of Reading, Whiteknights, PO Box 220, Reading, RG6 6AX (UK)

Department of Physics, University of Bologna, Viale Berti Pichat 6/2, 40127 Bologna (Italy)

Klaus Fraedrich and Frank Lunkeit

Meteorologisches Institut, KlimaCampus, University of Hamburg, Grindelberg 5, 20144 Hamburg (Germany)

## Abstract

We present an extensive thermodynamic analysis of a hysteresis experiment performed on a simplified yet Earth-like climate model. We slowly vary the solar constant by 20% around the present value and detect that for a large range of values of the solar constant the realization of snowball or of regular climate conditions depends on the history of the system. Using recent results on the global climate thermodynamics, we show that the two regimes feature radically different properties. The efficiency of the climate machine monotonically increases with decreasing solar constant in present climate conditions, whereas the opposite takes place in snowball conditions. Instead, entropy production is monotonically increasing with the solar constant in both branches of climate conditions, and its value is about four times larger in the warm branch than in the corresponding cold state. Finally, the degree of irreversibility of the system, measured as the fraction of excess entropy production due to irreversible heat transport processes, is much higher in the warm climate conditions, with an explosive growth in the upper range of the considered values of solar constants. Whereas in the cold climate regime a dominating role is played by changes in the meridional albedo contrast, in the warm climate regime changes in the intensity of latent heat fluxes are crucial for determining the observed properties. This substantiates the importance of addressing correctly the variations of the hydrological cycle in a changing climate. An interpretation of the climate transitions at the tipping points based upon macro-scale thermodynamic properties is also proposed. Our results support the adoption of a new generation of diagnostic tools based on the 2$^{nd}$ law of thermodynamics for auditing climate model and outline a set of parameterizations to be used in conceptual and intermediate complexity models or for the reconstruction of the past climate conditions.

# Introduction

The investigation of the global structural properties plays a central role for the provision of a unifying picture of the climate system on a large variety of time and space scales as, envisioned by the scientific perspective beautifully presented in the landmark book by Saltzman (2002).

Many authors have approached this problem mainly from a "deterministic dynamical systems" point of view, with the ensuing successes due to the discovery of – possibly - general mathematical properties – see, e.g., the analysis of the stability of the thermohaline circulation (Rahmstorf 1995) - and the flaws due to the facts that the simplifications adopted in the heuristic derivation of the dynamical systems sometimes blur out the involved physical processes and hardly allow for an efficient representation of the actual fluctuations and of the thermodynamics of the system. The introduction of stochastic forcing provides a relatively simple but conceptually rich partial solution to some of these problems, with, anyway, the unavoidable limitation of the need for – an usually beyond reach - "closure theory" for the statistical properties of noise. See Fraedrich (1978) and Saltzman (2002) for an in-depth discussion of these issues in a rather general setting.

Another way of addressing the "big picture" of the climate system is to adopt a thermodynamic perspective. The analysis of the energy cycle of the atmosphere (Lorenz 1955, 1967) highlighted the concept of availability by showing that only a tiny part of the potential energy of the atmosphere can be converted to mechanical energy. The formalization of the concept of efficiency allowed for defining an effective climate machine, driven by the temperature difference between a warm and a cold thermal pool. Thus, the atmospheric and oceanic motions are at the same time the result of the mechanical work (then dissipated by in a turbulent cascade) produced by the engine, and are processes which re-equilibrate the energy balance of the climate system (Lorenz 1967; Stone 1978; Peixoto and Oort 1992). A rigorous definition of a Carnot engine–equivalent picture of the climate system has been recently provided (Johnson 1989; 2000) and further expanded. On a parallel line, a great deal of attention has been paid to the application of non-equilibrium thermodynamics to the climate system. The climate can be seen as a non-equilibrium system, which generates entropy by irreversible processes and keeps a steady state by balancing the input and output of energy and entropy with the surrounding environment. Peixoto and Oort (1992) provided a detailed theoretical presentation of the entropy production in the climate system and some reasonable estimates of its value are given, whereas in Ozawa et al. (2003) and Kleidon and Lorenz (2005) a more modern perspective is given. Moreover, following the stimulations by Johnson (1997), detailed analyses of spurious entropy generation in dynamical cores of climate models have been recently presented (Woollings and Thuburn 2006). Finally, in spite of its debated validity (Dewar 2005; Grinstein and Linsker 2007), an increasing popularity is enjoyed by methods related to the maximum entropy production principle (MEPP), first discussed by Paltridge (1975). MEPP proposes that an out-of-equilibrium nonlinear system adjusts in such a way to maximize the production of entropy (Martyushev and Seleznev 2006). First applications of the MEPP to GCMs have been provided by Kleidon et al (2003, 2006) and Fraedrich and Lunkeit (2008).

Recently, a line connecting the investigation of the climate as a thermal engine to the analysis of its entropy production has been drawn (Lucarini 2009). Within the same perspective, in this paper we aim at exploring a major paleoclimatic transition – namely the onset and decay of snowball conditions (Saltzman 2002). Following the early stimulation by Kirschvink (1992) and the subsequent indication of the possible occurrence of snowball conditions for our planet in the Neoproterozoic (Hoffman et al. 1998), a great deal of attention has been placed upon understanding the conditions under which snowball conditions could be realized and, subsequently, a reverse transition to (relatively) a ice-free planet could be accomplished (Hoffman and Schrag 2002). Simplified energy balance models have emphasized that the ice-albedo feedback (Budyko 1969, Sellers, 1969) and the water-vapour feedback (Fraedrich 1979) constitute the fundamental mechanisms in determining such catastrophic (Arnold 1992) climate shifts. With the fast development of computer modeling and the possibility to take into account an increasing body of paleogeographic information, the simulation of snowball conditions has been extensively discussed in the climate modeling literature (see, e.g. Donnadieu et al. 2004). In particular, it has been guessed that, within a certain parametric range, the same irradiance can support both the ice-covered and the ice-free conditions (Pierrehumbert 2004), i.e. the climate system features bistability. As it is thought that the recovery from the guessed ice-covered state could have been provoked by a massive increase in $CO_2$ concentration, the impact of changes in atmospheric composition on deglaciation processes has been extensively studied. Nevertheless, when paleo-conditions are considered, recent simulations (Pierrehumbert 2005) have found unsatisfactory agreement with estimates obtained with simplified models (Caldeira and Kastings 1992). More recent studies, performed with a state-of-the-art fully coupled climate model with present-day boundary conditions, have shown that recovery from ice-covered state is indeed possible with injection of large amounts of CO2 in the atmosphere (Marotzke and Botzet 2007; Voigt and Marotzke 2009). These studies have also emphasized the role of ocean transports in determining the stability properties of the system.

In this paper, rather than trying to answer the specific question on whether the planet Earth underwent an actual cycle of global glaciation/deglaciation, we wish to study, using a simplified yet Earth-like climate model (Planet Simulator, Fraedrich et al., 2005a,b), how the macro-scale thermodynamics of the climate system changes when parametrically controlling the value of the solar constant and exploring glaciation/deglaciation transitions. Following the theoretical framework described in Lucarini (2009), we compute the Carnot efficiency, the entropy production and the degree of irreversibility of the climate system for a large set of simulations representative of climate conditions realized for value of the solar constants ranging between 90% and 110% of the present value. Therefore, the goal we pursue is to propose a framework suitable for defining in a comprehensive way possibly new interpretations of large scale climate processes and of major climate transitions. This has significant relevance also in the context of the ever-increasing attention paid by the scientific community to the quest for reliable metrics to be used in the validation of climate models of various degrees of complexity, as discussed, e.g., in Held (2005), Lucarini et al. (2007a), Lucarini (2008). The paper is organized as follows. In Section 2 we recapitulate the theoretical framework considered in this analysis. In Section 3 we overview the main features of the climate model, of the performed simulations, of the data analysis strategies, including some techniques able to circumvent some unavoidable model errors. In Section 4 we present our main results, in Section 5 our conclusions are drawn and perspective future works outlined.

## 2. Notes on the non-equilibrium thermodynamics of the Climate System

Let the total energy of the $\Omega$ domain of the climatic system be given by $E(\Omega) = P(\Omega) + K(\Omega)$, where $P$ represents the moist static potential energy, given by the thermal - inclusive of the contributions due to water phase transitions – and potential contributions, and $K$ is the total kinetic energy. The time derivative of the total kinetic and potential energy can be expressed as $\dot{K} = -\dot{D} + \dot{W}$, $\dot{P} = \Psi + \dot{D} - \dot{W}$, where we have dropped $\Omega$-dependence for convenience, $\dot{D}$ is the (positive definite) integrated dissipation, $\dot{W}$ is the instantaneous work performed by the system, $\Psi$ represents the heating due to convergence of heat fluxes (which can be split into the radiative, sensible, and latent heat components), such that $\dot{E} = \Psi$. We denote the total heating rate as $\dot{\Phi} = \Psi + \dot{D}$. If we are considering a non-equilibrium steady state system (NESS, see Gallavotti 2006), the quantities $E$, $P$, $K$ are stationary (in terms of statistical properties). Therefore, $\overline{\dot{E}} = \overline{\dot{P}} = \overline{\dot{K}} = 0$, where the upper bar indicates time averaging over a long time scale. At any instant, we can partition the domain $\Omega$ into $\Omega^+$ and $\Omega^-$, such that the intensive total heating rate $\dot{Q} = 1/\rho\left(\varepsilon^2 - \vec{\nabla} \cdot \vec{H}\right)$, given by the sum of viscous dissipation and convergence of the turbulent and radiative heat fluxes $\vec{H}$, is positive in $\Omega^+$ and negative in $\Omega^-$. Therefore:

$$\dot{P} + \dot{W} = \int_{\Omega^+} dV \rho \dot{Q}^+ + \int_{\Omega^-} dV \rho \dot{Q}^- = \dot{\Phi}^+ + \dot{\Phi}^- = \dot{\Phi} \qquad (1)$$

where the quantities $\dot{\Phi}^+$ and $\dot{\Phi}^-$ are positive and negative at all times, respectively. Since dissipation is positive definite, we obtain $-\overline{\dot{K}} + \overline{\dot{W}} = \overline{\dot{D}} = \overline{\dot{P}} + \overline{\dot{W}} = \overline{\dot{W}} = \overline{\dot{\Phi}^+} + \overline{\dot{\Phi}^-} > 0$.

Using the 2nd law of thermodynamics, and assuming local thermodynamic equilibrium – which applies pretty well everywhere except in the upper atmosphere, which has a negligible mass – we have locally $\dot{Q} = \dot{s}T$, with $\dot{s}$ the time derivative of the entropy density. We hereby neglect the contributions to entropy variations due to chemical mixing related to salinity fluxes in the ocean (Shimokawa and Ozawa 2001) and due to the mixing of water vapour in the atmosphere (Pauluis and Held 2002); as discussed in Lucarini (2009), both terms can be safely neglected. The derivative of the entropy of the system is:

$$\dot{S} = \int_{\Omega^+} dV \frac{\rho \dot{Q}^+}{T} + \int_{\Omega^-} dV \frac{\rho \dot{Q}^-}{T} = \int_{\Omega^+} dV \rho |\dot{s}^+| - \int_{\Omega^-} dV \rho |\dot{s}^-| = \dot{\Sigma}^+ + \dot{\Sigma}^- \qquad (2)$$

where at all times $\dot{\Sigma}^+ > 0$ and $\dot{\Sigma}^- < 0$. Since the system is at steady state $\overline{\dot{S}} = 0$, so that $\overline{\dot{\Sigma}^+} = -\overline{\dot{\Sigma}^-}$. Moreover, $2\overline{\dot{\Sigma}^+} = \overline{\int_\Omega dV \rho |\dot{s}|}$, so that $\overline{\dot{\Sigma}^+}$ measures the absolute value of the entropy fluctuations throughout the domain. Using the mean value theorem, we obtain that $\overline{\dot{\Phi}^+} = \overline{\dot{\Sigma}^+}\Theta^+$ and $\overline{\dot{\Phi}^-} = \overline{\dot{\Sigma}^-}\Theta^-$, where $\Theta^+$ ($\Theta^-$) is the time and space averaged value of the temperature where absorption (release) of heat occurs. Since $\left|\overline{\dot{\Sigma}^+}\right| = \left|\overline{\dot{\Sigma}^-}\right|$ and $\left|\overline{\dot{\Phi}^+}\right| > \left|\overline{\dot{\Phi}^-}\right|$ we derive that $\Theta^+ > \Theta^-$, *i.e.* absorption typically occurs at higher temperature than release of heat (Peixoto and Oort 1992; Johnson 1989; 2000). We then obtain that the climate system can be represented as a Carnot engine such that $\overline{W} = \eta \overline{\dot{\Phi}^+}$, where $\eta = (\Theta^+ - \Theta^-)/\Theta^+ = (\overline{\dot{\Phi}^+} + \overline{\dot{\Phi}^-})/\overline{\dot{\Phi}^+}$ can be rigorously defined as the equivalent Carnot efficiency. As shown in Lorenz (1955) - and clarified in Johnson (2000) -, the long term average of the work performed by the system is equal to the long term average of the generation of available potential energy, which can be interpreted as the portion of the total potential energy which is available for reversible conversion.

The 2$^{nd}$ law of thermodynamics states that the entropy variation of a system at temperature $T$ receiving an amount of heat $\delta Q$ is larger than or equal to $\delta Q/T$ (Landau and Lifschitz 1980). In this case we have:

$$\overline{\dot{S}_{in}(\Omega)} \geq \overline{\dot{S}_{min}(\Omega)} = \overline{\left(\frac{\int_\Omega dV \rho \dot{Q}}{\int_\Omega dV \rho T}\right)} = \overline{\left(\frac{\dot{\Phi}^+ + \dot{\Phi}^-}{\langle\Theta\rangle}\right)} \approx \frac{\overline{\dot{\Phi}^+ + \dot{\Phi}^-}}{\overline{\langle\Theta\rangle}} \approx \frac{\overline{\dot{\Phi}^+ + \dot{\Phi}^-}}{(\Theta^+ + \Theta^-)/2} = \frac{\overline{W}}{(\Theta^+ + \Theta^-)/2}, \quad (3)$$

where $\overline{\dot{S}_{in}(\Omega)}$ is the long-term average of the entropy production inside the system, $\overline{\dot{S}_{min}(\Omega)}$ is its lower bound, *i.e.* the minimal value of the entropy production compatible with the presence of a Lorenz energy cycle with intensity $\overline{W}$, $\langle\Theta\rangle$, and $\langle\Theta\rangle$ is the density-averaged temperature of the system. The approximation holds as long as we can neglect the impact of the time cross-correlation between the total net heat balance and the average temperature and we can assume that $\langle\Theta\rangle$ can be approximated by the mean of the two Carnot temperatures $\Theta^+$ and $\Theta^-$. We can explicitly write $\overline{\dot{S}_{min}(\Omega)}$ as:

$$\overline{\dot{S}_{min}(\Omega)} \approx \frac{\overline{W}}{(\Theta^+ + \Theta^-)/2} = \frac{\eta \overline{\dot{\Phi}^+}}{(\Theta^+ + \Theta^-)/2} = \eta \frac{\Theta^+}{(\Theta^+ + \Theta^-)/2} \overline{\dot{\Sigma}^+} = \eta \frac{1}{1-\eta/2} \overline{\dot{\Sigma}^+} \approx \eta \overline{\dot{\Sigma}^+} \quad (4)$$

where the last approximation holds as long as $\eta \ll 1$, which applies in the case of the climate system. Note that we are hereby excluding the - quantitatively dominant, but of dubious actual dynamical relevance, see Ozawa et al. (2003), Kleidon et al. (2003, 2006) - entropy production due to the degradation of solar UV radiation into terrestrial IR radiation. Therefore, $\eta$ sets also the scale relating the lower bound to the entropy production of the system $\overline{\dot{S}_{min}(\Omega)}$ – due to macroscopically irreversible processes - to the absolute value of the entropy fluctuations inside the system due to microscopically reversible heating or cooling processes. Note that if the system is isothermal and at equilibrium the internal entropy production is zero, since the efficiency $\eta$ is vanishing. We can express the average material entropy production as $\overline{\dot{S}_{in}} = \overline{\dot{S}_{min}} + \overline{\dot{S}_{exc}}$, where $\overline{\dot{S}_{exc}}$ is the excess of entropy production with respect to the minimum, which results from the heat transport downgradient the temperature field. We can define:

$$\alpha \approx \frac{\dot{S}_{exc}}{\dot{S}_{min}} \approx \frac{\overline{\int_\Omega dV \vec{H} \cdot \vec{\nabla}\left(\frac{1}{T}\right)}}{\overline{\dot{W}}/\langle\Theta\rangle} \geq 0 \qquad (5)$$

as a parameter of irreversibility of the system, which is zero if all the production of entropy is due to the – unavoidable - viscous dissipation of the mechanical energy. As $\overline{\dot{S}_{in}} \approx \eta \overline{\Sigma}^+ (1+\alpha)$, we have that entropy production is maximized if we have a joint optimization of heat transport and of production of mechanical work. Note that, if heat transport down-gradient the temperature field is very strong, the efficiency $\eta$ is small because the difference between the temperatures of the warm and the cold reservoirs is greatly reduced (the system is almost isothermal), whereas, if the transport is very weak, the factor $\alpha$ is small. This provides a new way of interpreting the controversial MEPP, and in particular clarifies why in Kleidon et al. (2003, 2006) the entropy production is maximized when the efficiency of the meridional heat transport is neither too high nor too low.

## 3.    Data and Methods

**3.1 Model and experimental design**

The Planet Simulator (PLASIM) is a climate model of intermediate complexity, freely available at http://www.mi.uni-hamburg.de/plasim (Fraedrich et al., 2005a,b). The primitive equations formulated for vorticity, divergence, temperature and the logarithm of surface pressure are solved via the spectral transform method (Orszag 1970; Eliasen et al. 1970). The parameterizations for unresolved processes consist of long (Sasamori 1968) and short (Lacis and Hansen 1974) wave radiation, interactive clouds (Slingo and Slingo 1991; Stephens 1978; Stephens et al. 1984), moist (Kuo 1965; 1974) and dry convection, large scale precipitation, boundary layer fluxes of latent and sensible heat and vertical and horizontal diffusion (Laursen and Eliasen 1989; Louis 1979; Louis et al. 1982; Roeckner et al. 1992). The land surface scheme uses five diffusive layers for the temperature and a bucket model for the soil hydrology. The model includes a 50m mixed-layer (swamp) ocean which includes a 0-dimensional thermodynamic sea ice model. In spite of their obvious limitations, as recently discussed by Danabasoglu and Gent (2008), slab ocean climate models are well suited for providing an accurate steady state climate response.

In a previous study (Fraedrich and Lunkeit 2008) PLASIM has been subjected to an analysis of the entropy budget and its sensitivity by directly computing the contribution from each individual process to the total entropy production. Compared to that study, one major modification of the model is made. Following Lucarini and Fraedrich (2009), the global atmospheric energy balance is greatly improved with respect to previous versions of the model by re-feeding the kinetic energy losses due to surface friction and horizontal and vertical momentum diffusion. See Becker (2003) for an in-depth discussion on these topics. PLASIM performs satisfactorily well, as the average energy bias is in all simulations smaller than $0.2\,Wm^{-2}$, which is about one order of magnitude smaller than most state-of-the art climate models (Lucarini and Ragone 2009). This is done locally by an instantaneous heating of the environmental air.

Here, PLASIM is used in the configuration with T21 horizontal resolution (approx. 5.6° x 5.6° on the corresponding Gaussian grid) with five sigma levels in the vertical. For the present study the usually prescribed oceanic heat transport is set to zero. A parameter sweep (or hysteresis) is performed by changing the solar constant. The experiment consists of 81 simulations: (i) From a present day value (1365 W/m$^2$) the solar constant is gradually decreased by steps of 0.5% of the present day value (i.e. 6.825W/m$^2$) to 1228.5 W/m$^2$ (-10%); (ii) then it is gradually increased to 1501.5 W/m$^2$ (+10% of the present day value); (ii) finally, it is gradually decreased again to the present day value. Each simulation (with a fixed solar constant) has a length of 100 years to ensure the system to archive equilibrium well before the end of the run. The next simulation (with modified solar constant) is started from the end of the previous one (the first present day integration is started from an initial state with an atmosphere at rest). The statistical properties of the observables considered in this study are computed on the last 30 years of the simulations in order to rule out the presence of transient effects. We consider, in addition to the global, time averaged surface temperature $\overline{T_s}$ the following bulk thermodynamic observables: $\overline{\dot{\Sigma}^+}$, $\overline{\dot{\Sigma}^-}$, $\overline{\dot{\Phi}^-}$, $\overline{\dot{\Phi}^+}$ (computed as described in Section 2) and $\overline{\dot{S}_{in}}$, computed as described in (Kunz et al. 2008; Ozawa et al. 2003).

**3.2 Accounting for spurious biases in the model diagnostics**

The theoretical framework proposed in section 2 needs some amendments in order to be applicable to actual data as it requires a rigorous agreement with the NESS hypothesis - namely, when first moments are considered, $\overline{\dot{E}} = \overline{\dot{P}} = \overline{\dot{K}} = \overline{\dot{S}} = 0$. If this condition is not obeyed, the definition of the efficiency $\eta$, which is central in our argumentation, becomes problematic. In spite to the improvements to the energy budget described above, PLASIM presents some modest biases in the global energy and entropy budgets, due basically to numerical errors affecting heating/cooling terms. This is a – relatively neglected - inadequacy common to all climate models, which has been the subject of a recent investigation (Lucarini and Ragone 2009) focusing on the energy budget of all the GCMs included in the 4th IPCC Assessment Report (IPCC 2007). We anticipate that PLASIM budgets are closer to zero than those of any state-of-the-art GCM. Moreover, the purely "adiabatic" numerical error in the entropy budget discussed thoroughly in (Johnson 2000) has been estimated at least one order of magnitude smaller than what discussed below, so that it will not be further considered.

Let us assume that the model errors are relatively small, so that $\overline{\dot{S}} = \overline{\dot{\Sigma}^+} + \overline{\dot{\Sigma}^-} = \Delta_S$, with $\Delta_S \ll \left|\overline{\dot{\Sigma}^+}\right|, \left|\overline{\dot{\Sigma}^-}\right|$, and that $\overline{\dot{E}} = \overline{\dot{P}} + \overline{\dot{K}} = \Delta_E$, with $\Delta_E \ll \left|\overline{\dot{\Phi}^+}\right|, \left|\overline{\dot{\Phi}^-}\right|$. If we assume that the kinetic energy budget is not biased – as actually is the case – we have $\overline{\dot{P}} = \Delta_E = \overline{\dot{\Phi}} - \overline{\dot{W}} = \overline{\dot{\Phi}^+} + \overline{\dot{\Phi}^-} - \overline{\dot{W}}$.

Therefore, $\overline{\dot{W}} = \overline{\dot{\Phi}^+} + \overline{\dot{\Phi}^-} - \Delta_E$, and we can simply define $\overline{\dot{W}} = \eta_E \overline{\dot{\Phi}^+}$ with $\eta_E = \left(\overline{\dot{\Phi}^+} + \overline{\dot{\Phi}^-} - \Delta_E\right)/\overline{\dot{\Phi}^+}$. Note that since $\Delta_E$ can be comparable to the algebraic sum of $\overline{\dot{\Phi}^+}$ and $\overline{\dot{\Phi}^-}$, neglecting the energy bias can cause notable errors in the definition of the efficiency and in the estimate of the average work $\overline{\dot{W}}$.

If, instead, we adopt straightforwardly the definition $\eta_\Theta = \left(\Theta^+ - \Theta^-\right)/\Theta^+$, and use the definitions $\overline{\dot{\Phi}^+} = \overline{\dot{\Sigma}^+}\Theta^+$ and $\overline{\dot{\Phi}^-} = \overline{\dot{\Sigma}^-}\Theta^-$, we obtain $\eta_\Theta = \left(\overline{\dot{\Phi}^+} - \overline{\dot{\Phi}^-}\overline{\dot{\Sigma}^+}/\overline{\dot{\Sigma}^-}\right)/\overline{\dot{\Phi}^+} = \left(\overline{\dot{\Phi}^+} + \overline{\dot{\Phi}^-} - \overline{\dot{\Phi}^-}\Delta_S/\overline{\dot{\Sigma}^-}\right)/\overline{\dot{\Phi}^+}$. Of course, if $\Delta_E = \Delta_S = 0$, $\eta_E = \eta_\Theta = \eta$. The 2$^{nd}$ law of thermodynamics imposes that the energy and entropy biases are related as follows: $\Delta_E \approx \langle\Theta\rangle\Delta_\Theta$, with a relative error of the order of $\approx \left(\Theta^+ - \Theta^-\right)/\langle\Theta\rangle \approx \eta$. Substituting in the expression of $\eta_\Theta$, we obtain:

$$\eta_\Theta \approx \left(\overline{\dot{\Phi}^+} + \overline{\dot{\Phi}^-} - \Delta_S \Theta^-/\langle\Theta\rangle\right)/\overline{\dot{\Phi}^+} \approx \left(\overline{\dot{\Phi}^+} + \overline{\dot{\Phi}^-} - \Delta_E\right)/\overline{\dot{\Phi}^+} = \eta_E \qquad (5)$$

Within a correction of the order of $\eta_\Theta^2$, the two definitions of efficiency are consistent. Therefore, the formula $\eta_\Theta = \left(\Theta^+ - \Theta^-\right)/\Theta^+$ can be used also in the case of a climate model featuring biases in the energy and entropy budgets without explicit need to diagnose $\Delta_E$ and $\Delta_\Theta$. For all the simulations described in Section 3a) we have verified that $\Delta_E$ and $\Delta_S$ are small and related up to a high degree of approximation as described above.

4.  **Results**

As a first step, we have analysed the dependence of the global average surface temperature $\bar{T}_S$ with respect to the solar constant $S_*$, in order to have a clear view of the large scale simulated climate conditions using as indicator the most commonly used benchmark variable in climate studies. For the present value $S_* = 1360\,Wm^{-2}$, we have as output $\bar{T}_S = 287.15\,K$, which, given the simplification involved in the PLASIM structure, is quite satisfactory. As shown in Fig. 1, for a large range of values of the solar constant we foresee two possible distinct climatic steady states, which can be first characterized by a large difference – of the order of 40 - 50 K - in the value of $\bar{T}_S$. Correspondingly, the system features two distinct branches, which correspond to the warm (W) – which the present climate belongs to – and the snowball (SB) climate conditions.

In the SB regime, the planet is characterized by a very cold climate corresponding to a greatly enhanced albedo due to the massive ice-cover. In our actual integration, we start from present climate conditions, decrease the value of $S_*$ and observe a sharp W→SB transition, corresponding to a bifurcation, occurring at $S_* \approx 1263\,Wm^{-2} \equiv S_*^{WSB}$. The value of the reduction of solar constant needed to induce the onset of snowball conditions basically agrees with what found by Poulsen and Jacob (2004) and by Voigt and Marotzke (2009), see also Weatherald and Manabe (1975, Fig. 5). The SB steady state is realized for values of the solar constant up to $S_* \approx 1433\,Wm^{-2} \equiv S_*^{SBW}$, where the reverse SB→W transition, with the corresponding bifurcation, is realized. Therefore, again in agreement with Voigt and Marotzke (2009), the present solar irradiance supports two rather distinct climatic states. These bifurcations correspond to the condition when the ice-albedo positive feedback is strong enough to destabilize the system. In Fig. 1 the left pointing triangle describes the transition from W→SB, the right pointing triangle indicates the SB→W transition. Moreover, the W branch is indicated with a red solid line, the SB branch in indicated with a blue solid line, whereas the transitions are depicted with black dashed lines. We will use this convention all throughout the paper. As expected in both SB and W conditions $\bar{T}_S$ monotonically increases with $S_*$, because the system has an input of larger amounts of energy. Far from the unstable region, we observe an extended quasi-linear regime with a constant sensitivity of $d\bar{T}_S/dS_* \approx 0.1 K/(Wm^{-2})$ in the W and $d\bar{T}_S/dS_* \approx 0.07 K/(Wm^{-2})$ in the SB branch, respectively. Instead, as typical when stability is lost, the sensitivity increases when the bifurcation is approached.

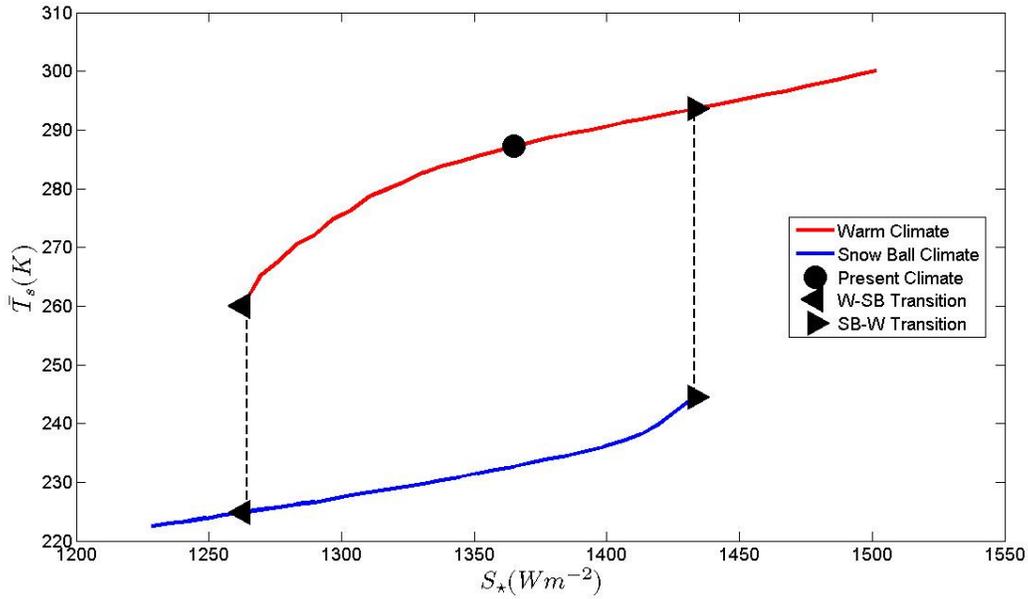

**Figure 1: Surface temperature T$_s$ against solar constant S$_*$. Details in the text.**

At this point a mathematical clarification is required. In simple models - see, e.g., Scott et al. (1999) – or, in general, in non-chaotic models – see, e.g., Lucarini et al. (2007b) - bistability is typically associated to the presence of two fixed points separated by an unstable equilibrium, and the loss of stability on an equilibrium is related to a bifurcation determining the change in the sign of one eigenvalue of the linearized problem. In the present case, instead, the two branches define the presence of two parametrically-modulated (by the changes in the value of the solar constant) families of disjoint strange attractors, as in each climate state the dynamics of the system is definitely chaotic, as shown by the fact that , e.g., the system features variability on all time scales. This corresponds, in Lorenz's terminology, to having intransitive climate conditions (Lorenz 1967, Peixoto and Oort 1992). The loss of stability realized in the W→SB and SB→W transitions is related to the catastrophic disappearance of one of the two strange attractors, resulting from complicated bifurcations. In this case, transitive climate conditions are realized. Near the transitions, the system features quasi-transitive climate conditions, where long transients can be observed.

**4.1 Global work**

We next investigate the time-averaged global work, computed as $\overline{\dot{W}} = \overline{\dot{\Phi}^+} + \overline{\dot{\Phi}^-} - \Delta_E$, which results into production of kinetic energy, eventually dissipated by viscous dissipation. Notably, - see Fig 2a - the work decreases with increasing solar constant in W climate conditions (with plateaus at both ends). This is agreement with the recent findings of Hernandez-Deckers and von Storch (2009), who observed in a full coupled GCM a decrease in the amount of kinetic energy production in a warmer climate with increased $CO_2$ concentrations. Note that smaller values of $\overline{\dot{W}}$ imply smaller values of dissipation, which in turn imply smaller intensity of surface winds. As opposed to that, in SB conditions the time-averaged global work increases with $S_*$, thus implying that a warmer climate is accompanied with enhanced generation of kinetic energy and stronger dissipation. This constitutes a very important difference between the W and SB climate conditions, which we will further explore below. As a further result of this discrepancy, we have that the jump in the value of $\overline{\dot{W}}$ in the W→SB transition is much larger than that in the SB→W transition.

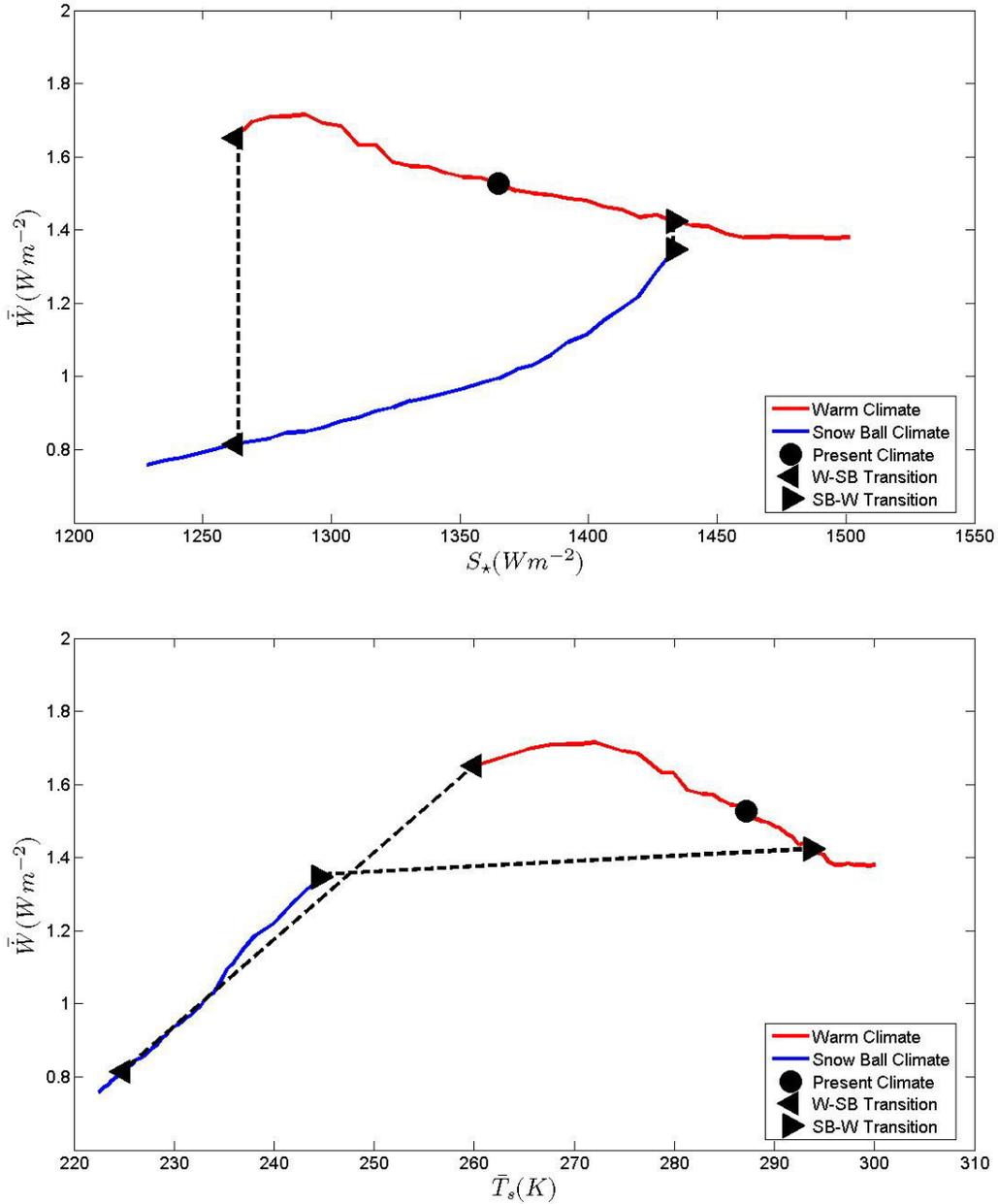

**Figure 2a-b: Time average of the global work (conversion from moist static energy to kinetic energy). In a), the work is plotted against the solar constant, in b) it is plotted against surface temperature**

Moreover, as for the surface temperature, in a relatively large parametric range $\overline{\dot{W}}$ is approximately linear with $S_*$. We can then robustly define the sensitivity of the kinetic energy production with respect to the solar constant, with resulting values of $\mathrm{d}\overline{\dot{W}}/\mathrm{d}S_* \approx -1.5 \cdot 10^{-3}$ in the W and $\mathrm{d}\overline{\dot{W}}/\mathrm{d}S_* \approx 1.5 \cdot 10^{-3}$ in the SB climate regime.

Since $\bar{T}_S$ is monotonically increasing with respect to $S_*$ in both the W and SB branches, we can plot $\bar{W}$ as a function of $\bar{T}_S$ (see Fig. 2b). $\bar{W}$ is almost perfectly linear with respect to $\bar{T}_S$ in the SB regime, whereas in the W regime linearity is restricted within 20 K around the present climate conditions. Such a graphical/functional representation could be especially useful for developing simplified parameterisations to be used in minimal models.

**4.2 Efficiency**

What is observed in Figs 2a-b is made clearer when considering the thermodynamic efficiency $\eta$ of the system. As thoroughly analysed in Section 3, the expression $\eta_\Theta$ is valid also in the presence of small energy and entropy biases, so we have adopted it in the actual computations of the efficiency in this work. Fig. 3a shows that the thermodynamic efficiency of the present climate is about 0.025 and all the analysed climate states considered in this work in this context feature an efficiency ranging between 0.01 and 0.045. Note that the range of variations of work (Figs. 2) is about half of that of $\eta$ because of the (partially) compensating variation of $\bar{\dot{\Phi}}^+$, which is higher in W than in SB conditions and increases monotonically with $\bar{T}_S$ and $S_*$.

The efficiency decreases with a marked linear behaviour with increasing solar constant in W conditions, whereas the behaviour is opposite in SB conditions. We can estimate the linear sensitivities as $d\eta/dS_* \approx 4\cdot 10^{-5} W^{-1} m^2$ and $d\eta/dS_* \approx -1.6\cdot 10^{-4} W^{-1} m^2$ for the SB and W climate regimes, respectively.

Such a discrepancy in the response of the fundamental property of the Carnot engine of the climate system defines, in terms of the 2$^{nd}$ law of thermodynamics, a crucial difference between the SB and W regimes. The fact the increases in $S_*$ improve the efficiency of the climate engine in the SB regime and decrease it in the W regime can be interpreted as follows.

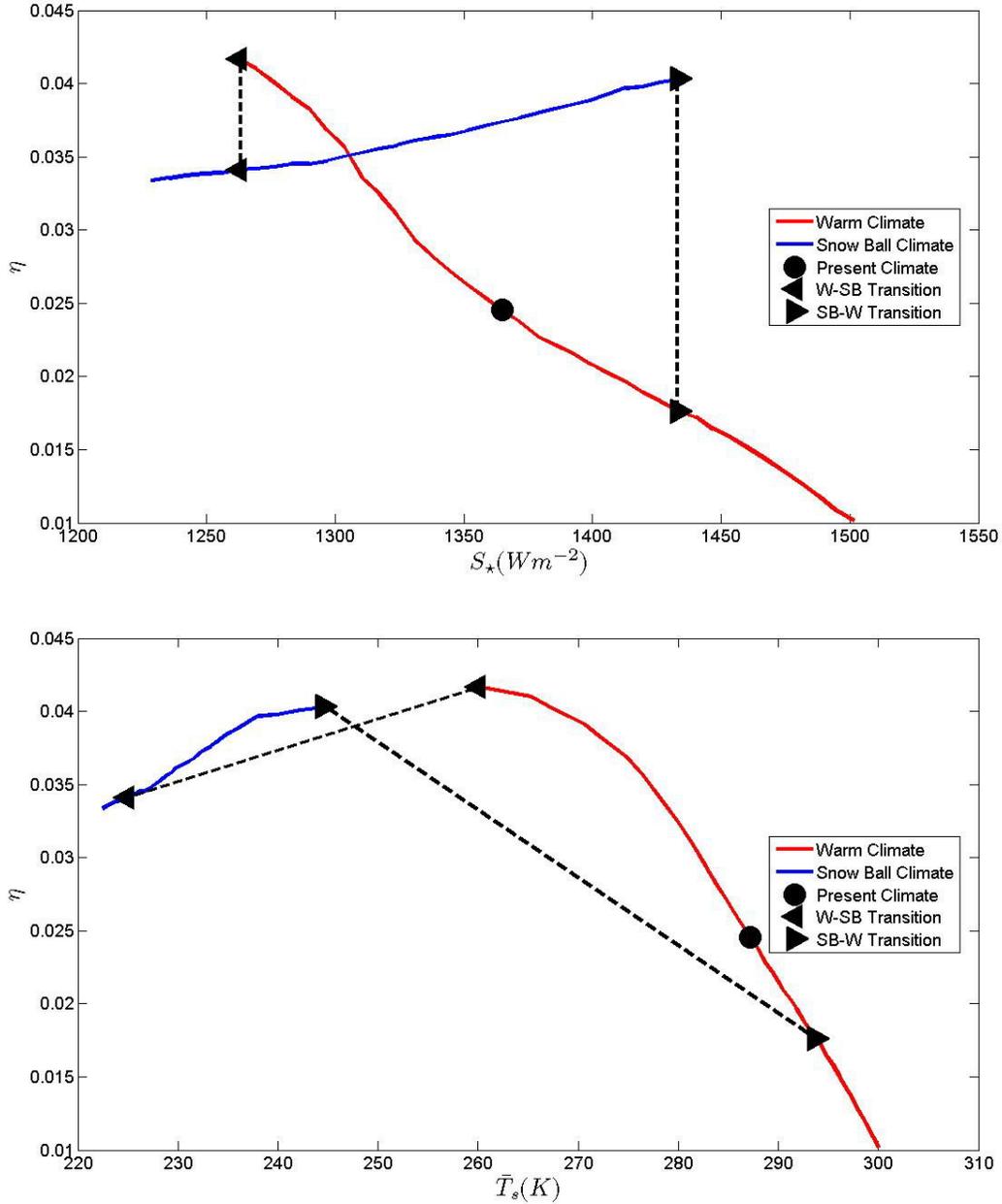

**Figure 3a-b:** Time average of the thermodynamic efficiency η (time and global averaged fraction of absorbed heart $Q_1$ converted into kinetic energy). In a), η is plotted against the solar constant, in b) it is plotted against surface temperature.

In the SB regime the main effect of increases in $S_*$ is favouring the melting of larger portions of the low-latitudes, with the resulting increase of the meridional temperature gradient due to the strong meridional albedo contrast, which leads to a more efficient – in terms of Carnot – climate machine. Instead, in the W regime, increases in the solar constant, thanks to the Clausius-Clapeyron effect (made intense by a warm background temperature) cause relevant enhancements in the efficacy of the turbulent latent heat fluxes, which tend to flatten the temperature gradients and impact negatively the thermodynamic efficiency of the system. In other terms: the warmer, the more isothermal in the W regime; the colder, the more isothermal in the SB regime (Fig 3b).

Moreover, we wish to underline that the crossing of the red and blue curves as observed in Fig. 3a implies that the efficiency decreases in both the W → SB and SB → W transitions. Since, as discussed in Section 3, a larger value of $\eta$ corresponds to a state more distant from equilibrium, this property agrees nicely with the fact that at the bifurcation the system make a transition to a more stable state.

Note also that, since the fractional variation of $\Theta_1$ as a function of the solar constant, being analogous of that of the surface temperature (see Fig. 1), is much smaller than that of $\eta$, Figs 3a-3b look almost identical to corresponding plots of $\Theta_1 - \Theta_2$ (not shown), with $\eta = 0.02$ corresponding roughly to $\Theta_1 - \Theta_2 = 5K$. Therefore, in the present climate we have $\Theta_1 - \Theta_2 \approx 6K$ and spanned over the whole range goes from $\Theta_1 - \Theta_2 \approx 3K$ to $\Theta_1 - \Theta_2 \approx 11K$. These numbers are always much smaller – by a factor of 5 or so – than the observed temperature difference between high and low latitudes which, therefore, are not reliable for providing an estimate of the climate efficiency.

**4.3 Entropy production**

We next analyse the material entropy production as described in Ozawa et al. (2003) and explicitly computed in Fraedrich and Lunkeit (2008) for PLASIM. Results are shown in Figs 4a-4b. The entropy production in both the W and SB branches monotonically increases with the solar constant, with an extended range of linear behaviour. In particular, we can define linear sensitivities whose values are $d\bar{\dot{S}}_{in}/dS_* \approx 4 \cdot 10^{-4} K^{-1}$ and $d\bar{\dot{S}}_{in}/dS_* \approx 1.5 \cdot 10^{-4} K^{-1}$ in the W and SB regime, respectively. For a given value of the solar constant, the entropy production in the W state is larger by a factor of about 4, and a wide gap exists between the range of the $\bar{\dot{S}}_{in}$ values in the W and in the SB regimes. Therefore, $\bar{\dot{S}}_{in}$ is a very well defined "state variable" of the climate system and seems to provide a more robust qualitative classification of the overall climate state than the surface temperature, and, in this sense, useful threshold values could be devised. When plotted against the surface temperature, the material entropy production is monotonically increasing with obviously a marked quasi-linear behaviour in both branches, so that simple parameterisations can be envisaged to be used in conceptual climate models.

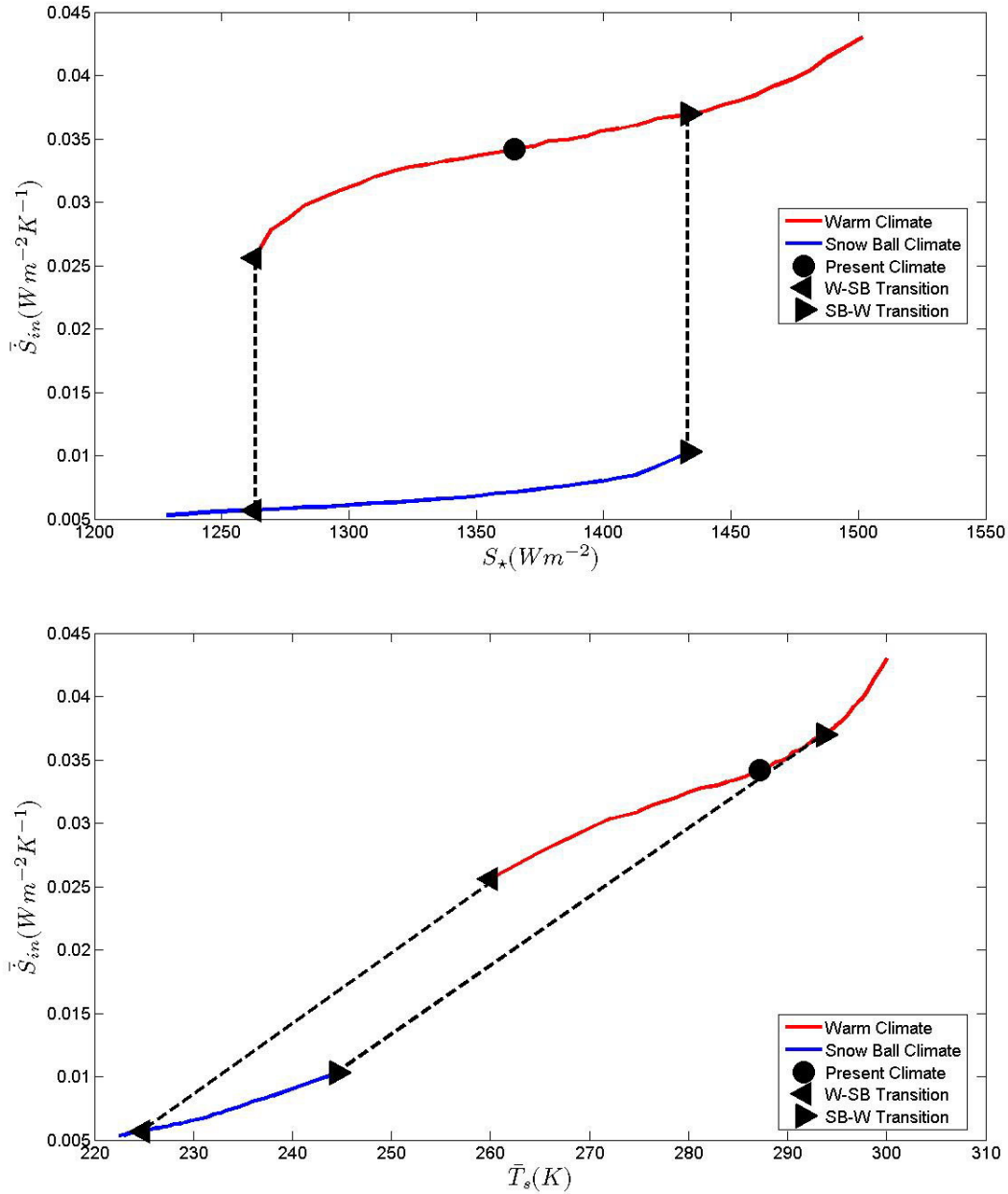

**Figure 4a-b: Material entropy production of the system, due to turbulent processes (radiation entropy is excluded). In a), the entropy production is plotted against the solar constant, in b) it is plotted against surface temperature, likewise the surface temperature (see Fig 1).**

**4.4 Irreversibility**

The last step in the analysis of the "2nd law" properties of the system to observe the behaviour of the factor $\alpha$, which, as described in Section 3, determines the degree of irreversibility of the system, The factor $\alpha = 0$ when no heat transport (via radiation or turbulent fluxes) occurs downgradient the temperature field. In this case, the only irreversible process is the – unavoidable - viscous dissipation of kinetic energy. Large values of $\alpha$ correspond to a "very irreversible" system. In the SB branch, $\alpha$ is small (<1), with a very weak quasi-linear dependence on $S_*$, such that we can estimate $\mathrm{d}\alpha/\mathrm{d}S_* \approx 10^{-3} W^{-1} m^2$. In the W branch, instead, the observed values of $\alpha$ are much higher and feature a fast increase with $S_*$, with a marked linear behaviour around the present climate conditions with $\mathrm{d}\alpha/\mathrm{d}S_* \approx 3 \cdot 10^{-2} W^{-1} m^2$. In the upper range of $S_*$ values, the increase becomes super-linear. The irreversibility parameter α is (weakly) linearly increasing with the surface temperature in the SB branch, whereas a much faster, almost explosive, behaviour is observed in the W branch. This clearly points at the role of the latent heat fluxes, whose strength depends exponentially on the temperature of the system, consistently – but from a different point of view – with what observed regarding the decrease in the efficiency of the system (Fig. 3).

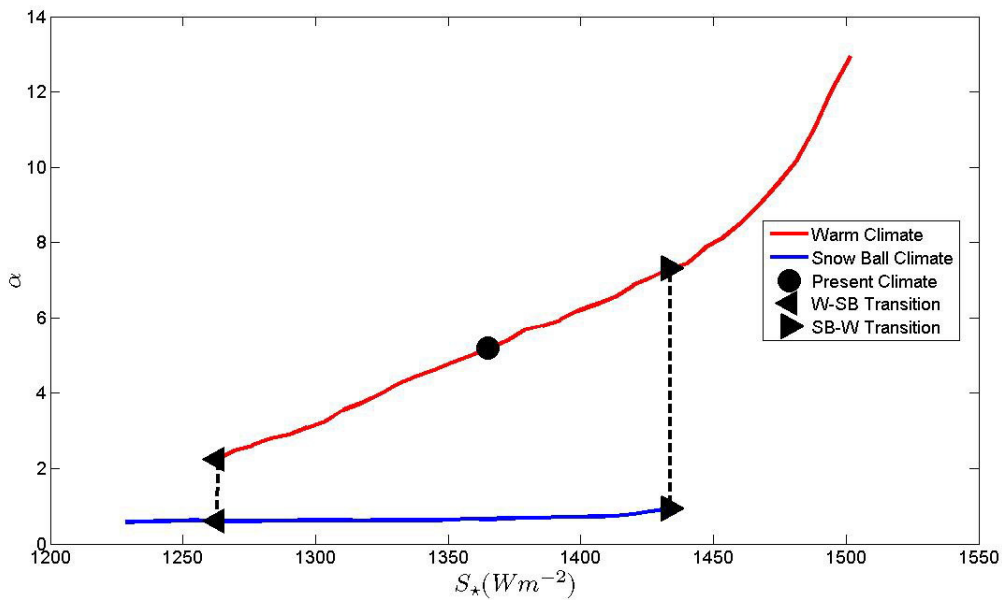

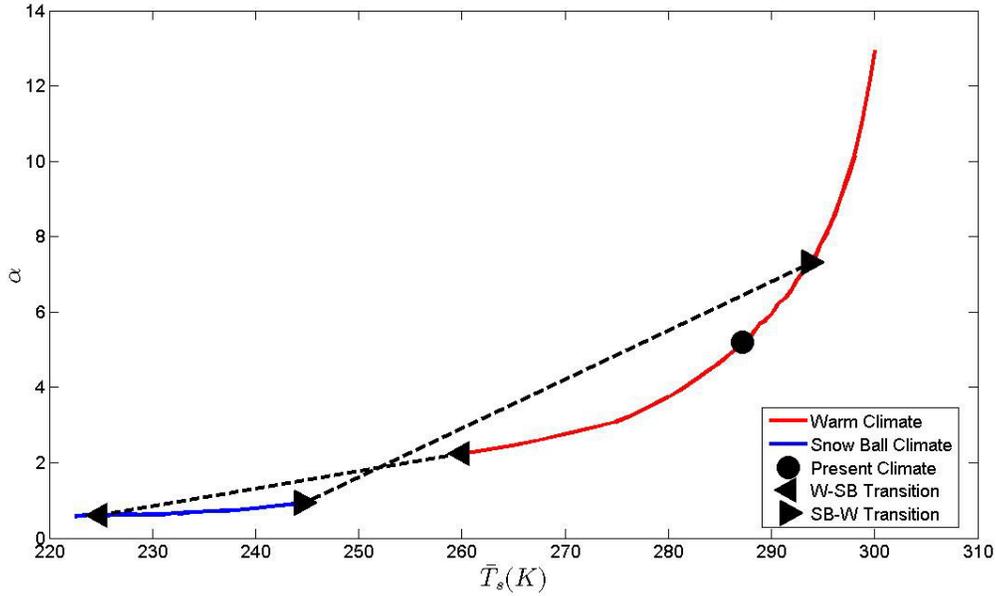

**Figure 5a-b: Degree of irreversibility of the system in terms of excess of entropy production or excess of exergy destruction.**

## 5. Conclusions

Since about twenty years nonlinear dynamics tools, which are based on the scaling properties of local time series, supplement the traditional climate data analysis. These methods provide information on predictability, long term memory and extremes. Here we introduce a set of novel diagnostic quantities, which are based on the laws of thermodynamics, to characterize global properties of climate system dynamics. These methods provide information on efficiency, entropy production, and irreversibility. Using this set of diagnostic tools derived from fundamental laws of physics based on the non-equilibrium thermodynamics, we have analyzed the warm (W) and snowball earth (SB) climate regimes of a general circulation model to identify some crucial differences between these regimes. From a basic point of view, this diagnostics characterizes features of the macro-state of the climate system. While the W and SB regimes agree, as expected, in featuring a positive response of surface temperature to increases in the solar constant, their behavior is opposite when looking at the dependence of the efficiency and of the irreversibility.

In the SB regime, increases in the solar constant lead to increases in the work performed by the system and with the ensuing dissipation (Fig. 2); this is due to an enhancement of the efficiency of the system (Fig. 3), made possible by strong meridional albedo gradients on the top of increases in the heat absorption. The material entropy production increases with the solar constant (Fig. 4), with a dominating contribution given by the unavoidable viscous-driven dissipation for all values considered of the solar constant. The portion of entropy production due to irreversible heat transport processes is almost constant, which implies that the material entropy production scales linearly with the total dissipation (and the total work) of the system.

In the W regime, higher values of the solar constant lead to decreases in the work performed by the system (Fig. 2), which, together with the increases in the heat absorption, leads to a dramatic decrease of the efficiency (Fig. 3). This corresponds to the fact that the system becomes almost isothermal for high values of the solar constant, or, more precisely, the (spatial) correlation between heating and temperature becomes rather weak. The ensuing decrease in the entropy production due to viscous dissipation is more than compensated by the large increase in the entropy production related to irreversible heat transport down-gradient the temperature field, deriving from the temperature-driven enhancement in the latent hear fluxes. Therefore, the material entropy production of the system is greatly enhanced by increases in the solar constant (Fig. 4), and the degree of irreversibility, measured by ratio between the excess of entropy production and the lower bound to the material entropy production determined by the $2^{nd}$ law of thermodynamics, increases very rapidly (Fig. 5). The fact that in the warm climate regime changes in the intensity of latent heat fluxes are crucial for determining macro-scale thermodynamic properties substantiates the importance of addressing correctly the variations of the hydrological cycle in a changing climate.

All of diagnostics considered, feature, in both the SB and W regimes, smooth dependence with respect to the solar constant with extended regions where linear fits provide rather good approximations. This suggests the possibility of developing parameterizations schemes (e.g.

variables expressed as functions of the surface temperature) to be adopted in conceptual and intermediate complexity models or for the reconstruction of the past climate conditions.

The transitions between the W and the SB regime (both ways) are characterized by a decrease in the efficiency of the system. Since a larger value of $\eta$ corresponds to a state more distant from equilibrium, this property agrees conceptually with the fact that, at the bifurcations, the system makes a transition from an unstable to a stable state. Moreover, since a larger efficiency is related to the presence of larger fluctuations of the climate state, macro-scale destabilizing mechanisms are more easily triggered.

Well within the parametric region where two distinct stationary states are detected, instead, if we accept the principle that states with higher entropy production are more probable (Kleidon and Lorenz 2005), we may conclude that it is in general more likely to find the system in the W rather than in the SB conditions. This stimulating hypothesis agrees with the climate conditions observed in the planet in the last hundreds of million of years.

Our analysis may provide tools for understanding, apart from the specific problem of snowball Earth conditions inception/decay, the processes involved in the determination of the so-called climate "tipping points", *i.e.* the conditions under which catastrophes changes may occur for small variations in the boundary conditions or in the internal parameters of the system. In this sense, we plan to investigate using the tools described here the impacts of very large changes in the $CO_2$ concentration and we want to explore whether new insights on the processes controlling the runaway greenhouse effect can be attained. Of course, in order to have a more realistic and quantitative picture of all the processes involved in the Earth system the adoption of more advanced modelling tools is required. In particular, a better representation of the ocean dynamics would constitute a crucial improvement, considering the ocean basically provides the bulk of the climate signal on multi-decadal time scales. The – computationally - problematic side of adopting a more realistic representation is not that only fully coupled atmosphere-ocean models are more demanding in terms of computer power *per se*, but also that, by possessing longer internal time scales, the number of years of simulation required for realizing a good representation of the steady state statistical properties is higher than in the case analysed here.

Nevertheless, lower-complexity models can be of extraordinary importance for devising *Gedankenexperimente* and for providing guidance in the definition of specific crucial experiments to be performed with more complex models.

Our results support the adoption of a new generation of diagnostic tools for auditing climate models of any level of complexity (Held 2005, Lucarini et al. 2007a, Lucarini 2008). We maintain that the adoption of metrics of validation based on the $2^{nd}$ law of thermodynamics will contribute to improving dramatically the ability of current climate models in representing the global thermodynamical properties of the climate system. Future work in this direction includes the analysis of the impact of $CO_2$ increase of the thermodynamic properties of the system, and the computation of the entropy production in state-of-the-art climate models, to be estimated, following the suggestion by Ozawa et al. (2003), by using radiation budgets at the top of the atmosphere.

**Acknowledgement**

We appreciate referees' comments that helped improving the manuscript, Dr X. Zhu for advising us on recent modelling advances, and the KlimaCampus (Hamburg) for support (V.L.).